\documentclass[aps,pra,twocolumn,groupedaddress,superscriptaddress,showpacs]{revtex4-1}

\usepackage[english]{babel}
\usepackage{graphicx,psfrag,color,hyperref}
\usepackage{amssymb,latexsym,mathrsfs,amsmath,amsthm}
\usepackage{subfigure}

\usepackage[utf8]{inputenc}

\begin{document}
\title{Quantum Zeno Dynamics through stochastic protocols}

\author{Matthias M. Müller}
\affiliation{\mbox{QSTAR, LENS, and Department of Physics and Astronomy, University of Florence}, via G. Sansone 1, I-50019 Sesto Fiorentino, Italy}

\author{Stefano Gherardini}
\affiliation{\mbox{QSTAR, LENS, and Department of Physics and Astronomy, University of Florence}, via G. Sansone 1, I-50019 Sesto Fiorentino, Italy}
\affiliation{\mbox{INFN and Department of Information Engineering, University of Florence,} via S. Marta 3, I-50139 Florence, Italy}

\author{Filippo Caruso}
\affiliation{\mbox{QSTAR, LENS, and Department of Physics and Astronomy, University of Florence}, via G. Sansone 1, I-50019 Sesto Fiorentino, Italy}

\begin{abstract}
Quantum Zeno Dynamics is the phenomenon that the observation or strong driving of a quantum system can freeze its dynamics to a subspace, effectively truncating the Hilbert space of the system. It represents the quantum version of the famous flying arrow Zeno paradox.
Here, we study how temporal stochasticity in the system observation (or driving) affects the survival probability of the system in the subspace. In particular, we introduce a strong and a weak Zeno regime for which we quantify the confinement by providing an analytical expression for this survival probability. We investigate several dissipative and coherent protocols to confine the dynamics, and show that they can be successfully adapted to the stochastic version. In the weak Zeno regime the dynamics within the subspace effectively acts as an additional source of stochasticity in the confinement protocol.
Our analytical predictions are numerically tested and verified on a paradigmatic spin chain. As practical implications different coherent and dissipative confinement protocols allow to choose a trade-off between a probabilistic scheme with high fidelity (compared to perfect subspace dynamics) and a deterministic one with a slightly lower fidelity, which is a step towards better control in future quantum technologies.
\end{abstract}

\def\be{\begin{equation}}
\def\ee{\end{equation}}
\def\bea{\begin{eqnarray}}
\def\eea{\end{eqnarray}}

\flushbottom
\maketitle
\thispagestyle{empty}

\section{Introduction}
The dynamics of a quantum system under constant observation may become completely frozen: this phenomenon is called quantum Zeno effect (QZE)~\cite{Misra1}. At the very heart of quantum Zeno phenomena is the quantum mechanical concept of measurement back action, i.e. the ability to drive a given quantum state along specific quantum paths by measuring the system, as first observed by von Neumann \cite{Neumann1}. In the case of very frequent measurements, the system is continuously projected back to its initial state, and the back action confines the system dynamics within the Hilbert subspace defined by the measurement operator. Experimentally it has been observed, among others, in terms of ions \cite{Itano1}, polarized photons \cite{Kwiat:1995}, and cold atoms \cite{Fischer:2001}.

In a more general context, if the projections are related to a multi-dimensional Hilbert subspace, one deals with so-called quantum Zeno dynamics (QZD), where the system dynamics remains confined in the measurement subspace, without necessarily remaining constrained to its initial state \cite{Pascazio1,Pascazio2}. It has been first demonstrated in an experiment with a rubidium Bose-Einstein condensate in a five-level Hilbert space \cite{Schafer1}, and later in a multi-level Rydberg state structure \cite{Signoles1}.
Moreover, also a quantum quasi-Zeno dynamics regime has been introduced, where the system may leave the subspace between two measurements but returns to the subspace before the next measurement occurs \cite{Elliott1}.
On one side, these phenomena have foundational implications about the nature of quantum measurements, since they are a physical consequence of the statistical indistinguishability of neighboring quantum states in the Hilbert space \cite{Wootters,Smerzi1}. In this regard, recently the realizability of QZD has been investigated also in the case where the evolution of the quantum system between two consecutive measurements is affected by non-Markovian noise \cite{Zhang1}. On the other side, they become increasingly relevant also from the practical point of view, for example for robust quantum information processing, where entangled states may be protected from decoherence by means of projective measurements \cite{Plastina2008}. In quantum computation, moreover, the projection onto an arbitrary symmetric subspace of the whole Hilbert space has allowed the creation of decoherence-free subspaces~\cite{Lidar1}, that are invariant with respect to the non-unitary part of the dynamics. Recently, it has also been proved that the evolution of physical observables can be restrained by frequent measurements even while the quantum state changes randomly in time \cite{Wang1}. Additionally, Zeno phenomena can be achieved by means of strong dissipative processes, modeling the unavoidable interaction of a quantum system with the external environment. The measurement can then be understood to be the randomly occuring quantum jumps~\cite{Plenio} describing the interaction with the environment.
In this context, the assumption to take the time among each measurement cannot be sustained anymore. It may happen, indeed, that the time interval between two consecutive measurements is randomly varied, such that the system undergoes stochastic quantum Zeno dynamics (SQZD). This concept has been introduced recently for one-dimensional projective measurements as stochastic quantum Zeno effect (SQZE)~\cite{Shushin1}. Then, the survival probability to remain in the initial state becomes itself a random variable that takes on different values corresponding to different realisations of the measurements sequence. Exploiting the theory of large deviations (LD) \cite{Ellis1,Dembo1,Touchette1}, the stochastic behaviour of the survival probability can be quantified and characterized in terms of the probability distribution of the time interval between two consecutive measurements \cite{Gherardini1}. In particular, it has been proved by LD theory that the survival probability for an increasing number of measurements converges to its most probable value, i.e. the typical value for a single realisation of the stochastic process. It has allowed, for instance, to verify the ergodic hypothesis for a randomly perturbed quantum system \cite{Gherardini2}.

\begin{figure}[t]
\centering
 \includegraphics[width=\linewidth]{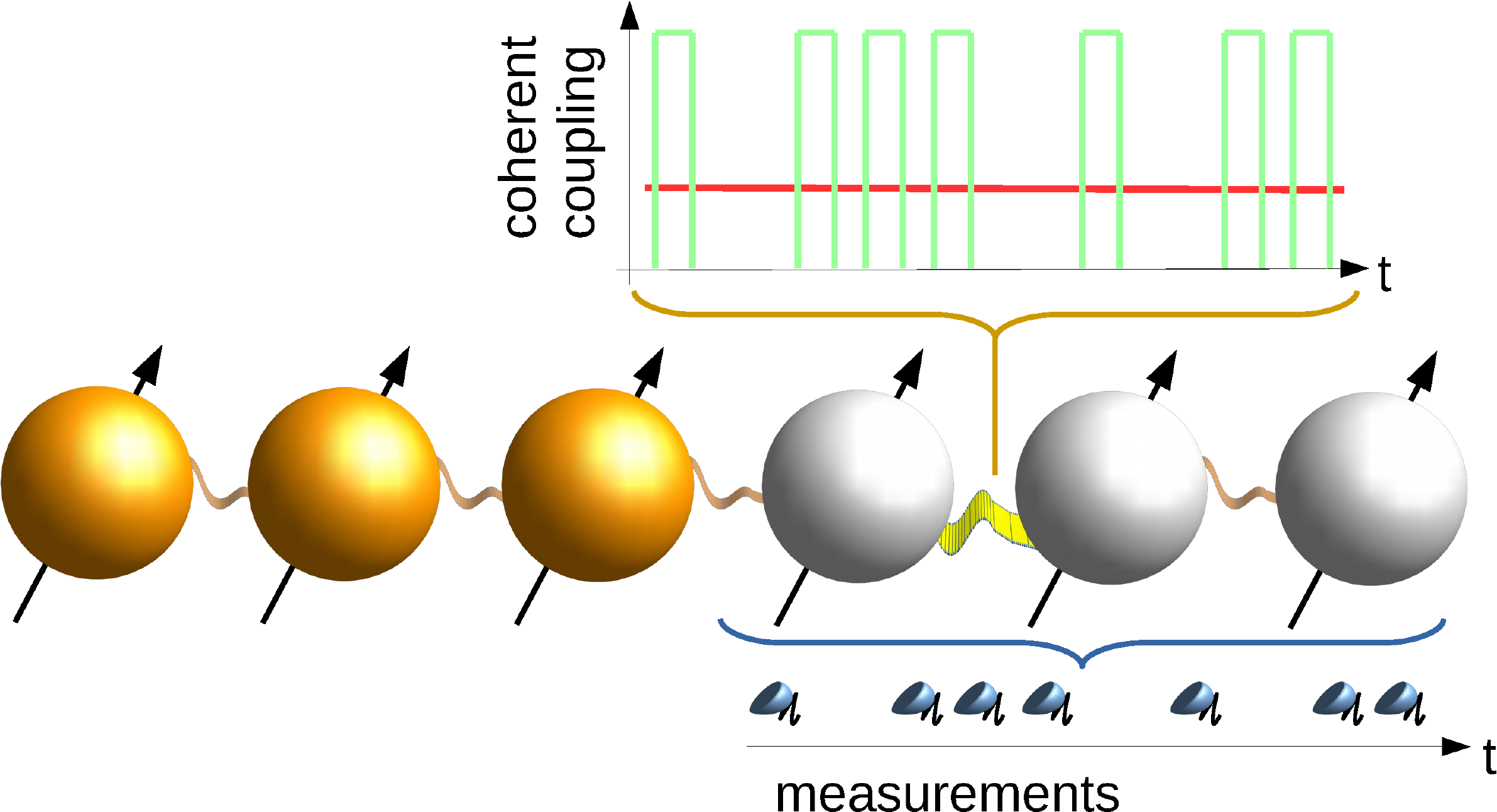}
 \caption{\textbf{Pictorial representation of the observation protocols for stochastic quantum Zeno dynamics.} A subsystem (orange, left) exhibits quantum Zeno dynamics when decoupled from the rest of the system by frequent measurements of the leakage from the subspace or by a strong coherent coupling effectively locking the dynamics of the border site. This coherent coupling can be continuous (red) or pulsed (green). The measurements (blue ``detectors'') as well as the coupling pulses can be spaced randomly thus making the leakage stochastic.}
 \label{fig:overview}
\end{figure}

Here, we investigate how the stochasticity in the time intervals between a series of projective measurements modifies the probability of a quantum system to be confined in an arbitrary Hilbert subspace. We generalize the LD formalism to SQZD in a regime where the dynamics within the subspace play a role. Moreover,
since both theoretically \cite{Pascazio1} and experimentally \cite{Schafer1} it has been demonstrated that QZD evolutions can be equivalently achieved not only by frequent projective measurements, but also by strong continuous coupling or fast coherent pulses, we study the accessibility to quantum Zeno dynamics if stochastic coherent or dissipative protocols are taken into account, as shown in Fig.~\ref{fig:overview}.
Our studies, thus, will provide a new tool in quantum information processing and quantum computation not only for controlling the amount of quantum coherence with Zeno-protection protocols, but also to design engineered quantum paths within the system Hilbert space along which the system dynamics is externally driven.  

\section{Model and theoretical results}
\subsection{Quantum Zeno dynamics and Stochasticity}
Let us consider a quantum mechanical system associated to a finite-dimensional Hilbert space $\mathcal{H}$, and a projector $\Pi$ onto the subspace $\mathcal{H}_{\Pi}\equiv \Pi\mathcal{H}$.
The scope of Zeno protocols is to constrain the dynamics of the system to remain within the subspace, which is, thus, called Zeno subspace. Its perfect implementation forbids the system to go beyond the Zeno subspace, such that, mathematically, the system dynamics is described exclusively by the projected Hamiltonian (or Zeno Hamiltonian) $\Pi H\Pi$, where $H$ is the full Hamiltonian acting on the Hilbert space $\mathcal{H}$. As a consequence, the evolution is determined by $U^{(\Pi)}(t)=\hat{T}\exp\left(-i\int_0^t \Pi H(t')\Pi dt'\right)$, where $\hat{T}$ is the time ordering operator. Hence,
$\rho^{(\Pi)}(t)=U^{(\Pi)}(t)\rho_0\big( U^{(\Pi)}(t)\big)^\dagger$ is the density matrix describing the system state within the Zeno subspace, where $\rho_0$ is the initial state.

For quantum Zeno dynamics, the standard observation protocols are given by applying a sequence of repeated projective measurements, separated by constant small intervals of free evolution of the system (with unitary dynamics), or by coherent dynamical couplings to another system playing the role of the measurement~\cite{Pascazio1,Pascazio2}. In the former case, the quantum state is projected onto the multidimensional subspace $\mathcal{H}_{\Pi}$ by the measurement operator $\Pi$, that is usually not commuting with the system Hamiltonian $H$. In the latter case, in the strong coupling limit, a dynamical super-selection arises that splits the Hilbert space into the eigenspaces of the coupling Hamiltonian.
Furthermore, the stochasticity introduces a possibility to engineer the dynamics by varying the underlying probability density function $p(\mu)$. This will pave the way towards exploring the whole Hilbert space of a quantum system, by engineering also the measurement operator dynamically thus slowly moving the population from one portion of the Hilbert space to another.
 We first describe a protocol based on projective measurements separated by random time intervals and then proceed to schemes with coherent driving.

\subsection{Stochastic projective measurements protocol}
Let us consider the dynamical evolution of a quantum system in the Hilbert space $\mathcal{H}$ if subjected to $m$ projective measurements separated by random time intervals $\mu_j$, $j=1,\dots,m$. The $\mu_{j}$'s are assumed to be independent and identically distributed random variables with probability density function $p(\mu)$. An arbitrary sequence of time-disordered measurements, thus, is characterized by a fixed number $m$ of measurements within a total time $t_{m}\equiv \sum_{j=1}^{m}\mu_j$ depending on the realisation of the $\mu_j$'s. Each realisation of the sequence $\{\mu_j\}\equiv\{\mu_j;~j=1,\ldots,m\}$ leads to a different system evolution, which is modelled by a density matrix $\rho(t)$. In particular, the unnormalised density matrix after $m$ measurements is given by $R_{m}(\{\mu_j\})\rho_{0}R_{m}^{\dagger}(\{\mu_j\})$, where we have defined the super-operator $R_m(\{\mu_j\})\equiv\prod_{j=1}^{m}\Pi U_{j}\Pi$, with $U_j$ the unitary evolution between measurement $j-1$ and $j$.
Note that from now on we assume $H$ to be constant to simplify the equations and thus have  $U_{j}\equiv\exp\left(-iH\mu_j\right)$. However, the qualitative results hold also for time-dependent Hamiltonians.
Consequently, the (survival) probability that the system belongs to the Zeno subspace $\mathcal{H}_{\Pi}$ at time $t_{m}$ is defined as
\be\label{Eq:surv_prob}
\mathcal{P}_m(\{\mu_j\})\equiv\rm{Tr}\left[R_{m}(\{\mu_j\})\rho_{0}R_{m}^{\dagger}(\{\mu_j\})\right],
\ee
which depends on the Hamiltonian $H$, the initial density matrix $\rho_{0}$, and also on the probability density function $p(\mu)$. Accordingly, the normalised density matrix at the end of the observation protocol is $\rho_m^{(p.m.)}(\{\mu_j\}) = \left[R_{m}(\{\mu_j\})\rho_{0}R_{m}^{\dagger}(\{\mu_j\})
\right]/\mathcal{P}_{m}(\{\mu_j\})$, where (p.m.) stands for \textit{projective measurements}.

The probability to find the system in the Zeno subspace at the $j-$th measurement will be denoted as $q_j(\mu_j)$, $j = 1,\ldots,m$.
Accordingly, the survival probability, i.e. the probability $\mathcal{P}_{m}(\{\mu_{j}\})\equiv\text{Prob}\left(\rho_{m}^{(p.m.)}\in\mathcal{H}_{\Pi}\right)$ that the system belongs to $\mathcal{H}_{\Pi}$ after $m$ projective measurements, is
$\mathcal{P}_{m}(\{\mu_{j}\})=\prod_{j=1}^{m}q_{j}(\mu_{j})$,
where $q_{j}(\mu_{j})\equiv\text{Tr}[\Pi\ U_{j}\Pi \rho^{(p.m.)}_{j-1}\Pi U_{j}^{\dagger}\ \Pi]$. For small $\mu_j$ the single survival probability $q_j(\mu_j)$ can be expanded \cite{Smerzi1}, i.e. $q_j(\mu_j)=1-\Delta^2_{\rho_{j-1}} H_{\Pi} \mu_j^2$, where $\Delta^2_{\rho_{j-1}} H_{\Pi}$ is the variance of $H_{\Pi}=H-\Pi H\Pi$ with respect to the state $\rho^{(p.m.)}_{j-1}$.
In the case of a unidimensional subspace given by the initial state, or more generally also when $t_{m}$ is small compared to the dynamics within the Zeno subspace, the survival probability reduces to $q_j(\mu_{j})\equiv q(\mu_{j})$: in the first case (unidimensional subspace) we have $q(\mu_j)=\text{Tr} [\Pi U_{j}\Pi \rho_{0}\Pi U_{j}^\dagger\Pi]$, in both cases $q(\mu_j)\approx 1-\Delta^2_{\rho_{0}} H_{\Pi} \mu_j^2$ (i.e. the variance is always calculated with respect to the initial state). With this simplification, the most probable value $\mathcal{P}^\star$ (for readability, we omit the index $m$) of the survival probability $\mathcal{P}_m(\{\mu_{j}\})$ is
\bea\label{eq:P_geom}
\mathcal{P}^\star=\prod_{\{\mu\}} q(\mu)^{mp(\mu)}
=\exp\left\{m\int_{\mu}d\mu p(\mu)\ln(q(\mu))\right\}.
\eea
Let us point out that, as has been recently demonstrated by LD theory~\cite{Gherardini1}, the survival probability of the sequence will converge to its most probable value for a large number of measurements $m$.
As shown in the appendix, by a Taylor expansion we find that under the condition
\be\label{eq:skewness_condition}
\int_{\mu}d\mu p(\mu)\mu^{3}\ll\frac{1}{mC},
\ee
where $C$ is a constant stemming from the remainder term of the expansion and depends on the specific system Hamiltonian $H$ and the initial state $\rho_{0}$, we can approximate the survival probability as
\be\label{eq:P_zeno2}
\mathcal{P}^\star\approx
\exp\left\{-m\Delta^2_{\rho_{0}}H_{\Pi} (1+\kappa)\overline{\mu}^2 \right\}.
\ee
Here, $\kappa=\frac{\Delta^{2}\mu}{\overline{\mu}^2}$ and $\overline{\mu}$ and $\Delta^{2}\mu$ are, respectively, the expectation value and variance of $p(\mu)$.
It is worth to note that the validity of Eq.~\eqref{eq:P_zeno2} (i.e. the condition Eq.~\eqref{eq:skewness_condition}) does not depend on the variance of the probability distribution $p(\mu)$, but on its degree of skewness.
This represents the first main result of this paper. As a matter of fact, Eq.~\eqref{eq:P_zeno2} generalizes the expression for the probability that the system belongs to the measurement subspace after $m$ random projective measurements beyond the standard Zeno regime~\cite{Smerzi1}. Consequently, we denote the inequality in Eq.~\eqref{eq:skewness_condition} as the \textit{weak Zeno limit}. In contrast the \textit{strong Zeno limit} requires $m\Delta^2 H_{\rho_0}(1+\kappa)\overline{\mu}^2\ll 1$, leading to
\be\label{eq:P-strictZeno3}
\mathcal{P}^\star\approx 
1 - m\Delta^{2}_{\rho_{0}}H_{\Pi}(1+\kappa)\overline{\mu}^{2},
\ee
which for $\kappa=0$ (sequence of equally-distributed measurements) is the survival probability for standard quantum Zeno dynamics~\cite{Smerzi1}.

More generally, when the measurement projector has dimension greater than one and the dynamics within the subspace plays a role, the previous simplification $q_j(\mu_j) = q(\mu_j)$ cannot be made anymore. In detail, when we have $q_j(\mu_j) = 1-  \Delta^2_{\rho_{j-1}}H_{\Pi}\mu_j^2$, we cannot any longer approximate $\rho_{j-1}$ by $\rho_0$. Instead, the dynamics within the subspace has to be taken into account. However, we can make a different approximation, namely we can approximate the state of the system by $\rho^{(\Pi)}(t)$ (the dynamics for perfect Zeno confinement). As a consequence, we expand the survival probability $q_{j}(\mu_{j})$ for small enough $\mu_j$ as $q_{j}(\mu_{j})=\tilde{q}(\mu_{j},c_{j}) = 1 - c_{j}^2\mu_{j}^2$, where $c_{j} = \Delta_{\rho^{(\Pi)}_{j-1}}H_{\Pi}$, and $\Delta^{2}_{\rho^{(\Pi)}_{j-1}}H_{\Pi}$ is the variance of $H_{\Pi}$ with respect to the density matrix $\rho_{j-1}^{(\Pi)}$. We introduce another (artificial) probability density function $\tilde{p}(c)$ for the coefficients $c_{j}$, that properly takes into account the average influence of the system dynamics on the leakage by requiring $\int\tilde p(c)c^{2}dc = \frac{1}{t_m}\int_{0}^{t_{m}}\Delta^{2}_{\rho^{(\Pi)}(t)}H_\Pi dt$. This allows us to write the survival probability as
\bea\label{eq:SQZD}
\mathcal{P}^{\star} = \prod_{\{c\}}\prod_{\{\mu\}}\left(\prod_{j=1}^{m}\tilde{q}(\mu_{j},c_{j})\right)^{p(\mu)\tilde{p}(c)}
\nonumber\\
=\exp\left\{m\int_{\mu,c}d\mu dc p(\mu)\tilde{p}(c)\ln(\tilde q(\mu,c))\right\}.
\eea
Moreover, still under the hypothesis that the quantum system is in the weak Zeno limit, we assume that $\Delta^{2}_{\rho^{(\Pi)}(t)}H_\Pi$ changes slowly compared to the measurement frequency. Hence, by making the approximation $\ln(\tilde q(\mu,c))\approx 1 - \tilde q(\mu,c)$, the integral in Eq.~\eqref{eq:SQZD} can be easily worked out:
\be\label{eq:SQZD2}
\mathcal{P}^\star\approx\exp\left\{ -\frac{m\overline{\mu}^{2}(1+\kappa)}{t_m}\int_{0}^{t_m}\Delta^{2}_{\rho^{(\Pi)}(t)}H_\Pi dt \right\}.
\ee
Eq.~\eqref{eq:SQZD2} is the generalization of Eq.~\eqref{eq:P_zeno2} for SQZD.

\subsection{Coherent Zeno protocols}
So far we have considered Zeno dynamics realised by instantaneous projective measurements. These measurements, however, are difficult to realise experimentally. Indeed, the duration of the measurement might be comparable to or even larger than the the time scale of the system dynamics. Alternatively, quantum Zeno dynamics can be achieved via coherent coupling~\cite{Pascazio2}: in particular \textit{continuous coupling} (c.c.) and \textit{pulsed coupling} (p.c.). An additional coupling Hamiltonian $gH_{c}$, acting on the complementary Zeno subspace $\mathcal{H}_{I-\Pi}$, is added to the system Hamiltonian $H$, and, in the limit of strong coupling strength $g$, different regions of the system Hilbert space can be dynamically disjointed. For the pulsed coupling protocol, the time intervals between two unitary kicks allow for the same stochasticity as the time-disordered measurements. 
These instantaneous rotations are given by $U^{(p.c.)}=\exp\left(-i H_c s\right)$. The effective time $s$ determines the rotation angle. This rotation angle is given by the pulse area of a coupling pulse in a finite time realisation. As in the case of quantum bang-bang controls for dynamical decoupling tasks \cite{Viola1}, we assume a finite pulse area, and practically arbitrary strong coupling kicks leading to practically instantaneous rotations. Similarly to the time-disordered sequence of projective measurements, also the Zeno protocol based on pulsed coupling is intrinsically stochastic, if the pulses are separated by the random time intervals $\mu_{j}$ sampled from $p(\mu)$. Accordingly, in order to make the results coming from the two coherent coupling schemes comparable, we require that on average the pulse area of the two coherent coupling protocols is the same. The survival probability is evaluated by computing $\mathcal{P} = \text{Tr}(\Pi\rho^{(c.c.)})$ or $\mathcal{P} = \text{Tr}(\Pi\rho^{(p.c.)})$, where $\rho^{(c.c.)}$ and $\rho^{(p.c.)}$ are the normalised density matrices of the system at the end, respectively, of the continuous and pulsed coupling Zeno protocol. As a matter of fact, a closed expression for the survival probability as a function of the coupling strength $g$ is not trivial to calculate.

However, we can derive the scaling of $\mathcal{P}$ with respect to $g$, taking into account, without loss of generality, the continuous coupling method. Hence, let us decompose the total Hamiltonian as $H_{tot} = \Pi H\Pi\otimes I + I\otimes (g H_{c} + (I-\Pi)H(I-\Pi)) + H_{int}$, where we have assumed that $H_{c}$ acts only outside the Zeno subspace, and $H_{int}$ is the interaction Hamiltonian term between the subspace and its complement. By transforming the total Hamiltonian in a basis where $H_{c}$ is diagonal, the coupling between the Zeno subspace and its complement is effectively a driving, that is off-resonant by a term proportional to $g$.
As a consequence, the confinement error $1-\mathcal{P}$ within the Zeno subspace scales as $||H_{int}||^2/g^2$.

This becomes clearer if we  consider the paradigmatic three level system given by the Hamiltonian $H_{tot} = \omega(|1\rangle\langle 2| + |2\rangle\langle 1|) + g(|2\rangle\langle 3| + |3\rangle\langle 2|)$
\cite{Pascazio2}. Specifically, the coupling rate to the upper level by the strength $g$ is playing the role of the measurement, and the Zeno subspace is assumed to be the state $|1\rangle$. The coupling Hamiltonian $H_{c}$, thus, is given by the term $g(|2\rangle\langle 3| + |3\rangle\langle 2|)$. We introduce a linear transformation $T$, which diagonalizes $H_{c}$ and makes the coupling diagonal. In the canonical matrix representation it reads
\be
T = \begin{pmatrix}
    1 & 0 & 0 \\
    0 & \frac{1}{\sqrt{2}} & \frac{1}{\sqrt{2}} \\
    0 & \frac{1}{\sqrt{2}} & -\frac{1}{\sqrt{2}}
    \end{pmatrix},
\ee
leading to the transformed Hamiltonian
\be
T^{\dagger}H T= \begin{pmatrix}
                 0 & \frac{\omega}{\sqrt{2}} & \frac{\omega}{\sqrt{2}} \\
                 \frac{\omega}{\sqrt{2}} & g & 0 \\
                 \frac{\omega}{\sqrt{2}} & 0 & -g
                \end{pmatrix}.
\ee
We can observe that, if the initial state of the system is taken in the Zeno subspace $\mathcal{H}_{\Pi}$, the coupling effectively makes it extremely difficult for the system dynamics to be transferred outside $\mathcal{H}_{\Pi}$, since the transition with respect to the rest of the Hilbert space (here, given by $\omega$) is moved out of resonance by a factor $g$. 
As a consequence the effective driving is reduced to $\omega^2/g^2$. When $g \rightarrow \infty$, we obtain an ideal confinement of the quantum system in the measurement subspace. This can be easily seen by solving the model, and computing the survival probability
\be
\mathcal{P}(t)=\left(1-\frac{2\omega^2}{\omega^2 + g^2}\sin^2\left(\frac{\sqrt{\omega^2 + g^2}\, t}{2}\right)\right)^2\,,
\ee
in the Zeno subspace~\cite{Pascazio2}. As a consequence the confinement error scales with one over the square of the coupling strength.

\section{Application and Numerics}
\subsection{Stochastic quantum Zeno Dynamics in spin chains}
The dynamics within the Zeno subspace can be characterized also by collective behaviours originating from inter-particle interactions. Hence, though the dynamics is confined to the Zeno subspace, the resulting dynamical complexity of the system can be exponentially larger\cite{Burgarth1}, hence increasing its controllability. We consider a chain of $N$ qubits whose dynamics is described by the Hamiltonian
\begin{eqnarray}
 H_N = \alpha \sum_{i=1}^N \sigma_z^{i} + \frac{\beta}{2}\sum_{i=1}^{N-1} \left(\sigma_x^i\sigma_x^{i+1} + \sigma_y^i\sigma_y^{i+1} \right)\,,
\end{eqnarray}
where $\sigma_z^i$ is the Pauli z-matrix acting on the $i$-th site, and $\sigma_{x/y}^i\sigma_{x/y}^{i+1}$ are the interaction terms, coupling spins $i$ and $i+1$ by the tensor product of the respective Pauli matrices \cite{Baxter1}. Moreover, $\alpha$ is an external (magnetic) field, while $\beta$ is the coupling strength of the interaction. We set $\alpha=\beta=2\pi\times 5\,$kHz. The measurement shall restrict the dynamics to excitations of the first $\lambda$ spins, defining, thus, a $2^{\lambda}$-dimensional subspace. If we measure the excitations outside this subspace, both the Hamiltonian evolution and negative measurement outcomes, giving the absence of population in the rest of the chain, preserve the number of excitations~\cite{Elliott1}. In the following, by choosing a proper initial state, we can limit ourselves to the single excitation sector, neglecting states with more than one excited spin, and consider only pure states of the form $|\psi(t)\rangle=\sum_{i=1}^{N} c_i(t)|1_i\rangle$, where $|1_i\rangle=|0..010..0\rangle$ is the state with one excitation at site $i$. Accordingly, the initial states will be chosen with $c_i(0)=0$ for $i>\lambda$. Hence, the probability to find the system in the subspace at the $j$-th measurement is
\begin{equation}
 q_j(\mu_j)= 1 - \beta^2 \mu_j^2 |c_{\lambda}(t_{j-1})|^2\,,
\end{equation}
where $\beta^2|c_{\lambda}(t_{j-1})|^2 = \Delta _{|\psi_{j-1}\rangle}^2 H_{\Pi}$ with the variance $\Delta _{|\psi_{j-1}\rangle}^2 H_{\Pi}$ calculated with respect to the previous state $|\psi_{j-1}\rangle$ of the system. Equivalently, the probability $q_j(\mu_j)$ could be computed directly by observing the population of the state $|1_{\lambda+1}\rangle$ at time $t_j$, which to second order corresponds to the leakage out of the subspace during the time interval $\mu_j$ when $|\psi(t_{j-1})\rangle$ lives in the subspace.

If we start from an eigenstate of the Zeno-Hamiltonian $\Pi H_N\Pi=H_{\lambda}$ (spin chain Hamiltonian with $\lambda$ spins) and we re-normalise the system state after every measurement, the coefficient $|c_{\lambda}(t)|^2=c_{\lambda}^2$ will be approximately constant. Thus, we have $q_j(\mu_j) = 1 - \beta^2 \mu_j^2 |c_{\lambda}(t_{j-1})|^2 = 1-\beta^2\mu_j^2 c_{\lambda}^2=q(\mu_j)$, i.e. the quantum mechanical probability of finding the system in the subspace upon measurements depends just on the length of the interval $\mu_{j}$. Accordingly, from Eq.~\eqref{eq:P_zeno2}, we have
\be\label{eq:P-const-exc}
\mathcal{P}^\star = \exp\left\{-\beta^{2}c_{\lambda}^{2}m \overline{\mu}^2(1+\kappa) \right\}.
\ee

However, in a more general case the time dependence of $|c_{\lambda}(t)|^2$ has to be taken into account. We can calculate $\mathcal{P}^{\star}$ either numerically by simulating the measurement sequence on the $N$ spin chain, or we make use of the approximation given by Eq.~\eqref{eq:SQZD2} for stochastic quantum Zeno dynamics, such that we can calculate $\mathcal{P}^{\star}$ from the dynamics of the subspace:
\be\label{eq:P-average-exc}
\mathcal{P}^\star\approx\exp\left\{ -\frac{m \beta^2\overline{\mu}^{2}(1+\kappa)}{t_m}\int_0^{t_m} |c_{\lambda}(t)|^2 dt \right\}.
\ee
Here, we present the numerical results for a chain of $N = 12$ spins and for two different initial states. All the results are evaluated for a bimodal distribution of the measurement intervals, namely $p(\mu) = p_{1}$ if $\mu=\mu^{(1)}$, or $p(\mu) = p_{2}$ if $\mu=\mu^{(2)}$. We first examine the behaviour of the survival probability when the system is subjected to the stochastic projective measurement protocol and examine two different initial states. Once, we prepare the state initially in an entangled W-state (i.e. a delocalized excitation) and then we consider an initial state where the excitation is localized in the left-most spin of the chain.

\subsubsection{W-state}
\begin{figure}[t]
\centering
\includegraphics[width=\linewidth]{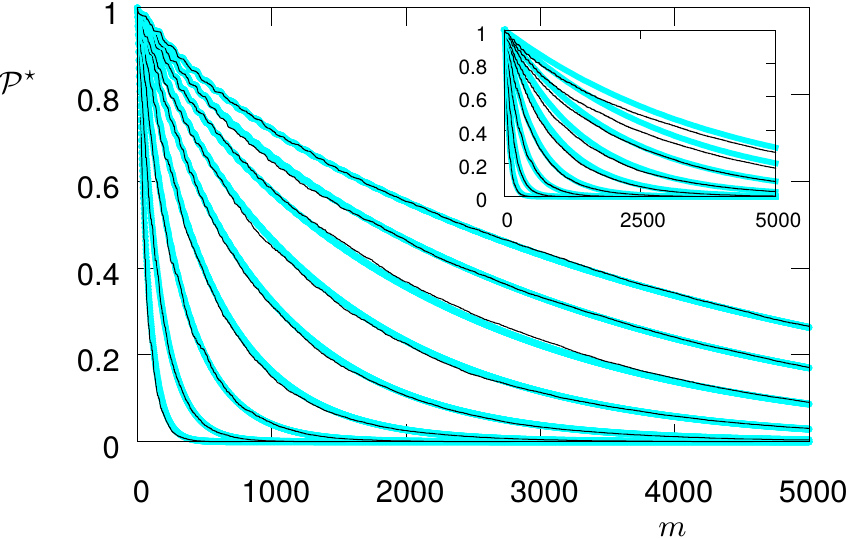}
\caption{\textbf{W-state.} One realisation of $\mathcal{P}$ (for each $\lambda=1,\dots,9$, from bottom to top) as a function of the number of measurements $m$ (black lines) compared to $\mathcal{P}^\star$ calculated by equation~\eqref{eq:P-average-exc} (cyan lines). Inset -- The same realisations (black lines) compared to $\mathcal{P}^\star$ calculated by equation~\eqref{eq:P-const-exc} (cyan lines). The probability density function is bimodal with $p_1 = p_2 = 0.5$, $\mu^{(1)}=1\,\mathrm{\mu s}$, and $\mu^{(2)}=5\,\mathrm{\mu s}$.}
\label{fig:W-P-exc}
\end{figure}

We first prepare the quantum system in the entangled state
\begin{equation}\label{27}
|\psi_{\lambda}(t)\rangle=\frac{1}{\sqrt{\lambda}}\sum_{i=1}^{\lambda} |1_i\rangle\,.
\end{equation}
Fig.~\ref{fig:W-P-exc} shows the survival probability, obtained by a numerical simulation of a random measurement sequence for $\lambda=1,\dots, 9$ (bottom to top), as compared to Eq.~\eqref{eq:P-average-exc} and an excellent agreement is observed.
Although the initial state of Eq.~\eqref{27} is not an eigenstate of $H_{\lambda}$, the dynamics of the system will approximately converge to one as observed in the numerical simulation. We can, thus, compare the survival probability $\mathcal{P}$, as obtained by the numerical simulation, to $\mathcal{P}^\star$ computed from Eq.~\eqref{eq:P-const-exc}, where $|c_{\lambda}(t)|^2$ is assumed to be constant and whose value is determined by taking the respective eigenstate. The inset of Fig.~\ref{fig:W-P-exc}, indeed, shows the comparison between this analytical approximation and the numerical values. The agreement is better for small $\lambda$, where the discrepancy between the initial state and the eigenstate is small (in particular, for $\lambda=1,2$ the initial state is an eigenstate of the subspace Hamiltonian $H_{\lambda}$).

\subsubsection{Left-most qubit excited}
By starting from $|1_1\rangle$ the excitation will travel towards the edge of the subspace where it is reflected. Hence, apart from the spreading, the excitation will oscillate between the edge of the chain and the edge of the subspace with a velocity $v$ approximately given by the Lieb-Robinson bound \cite{LR1972}.
\begin{figure}[h!]
\centering
\includegraphics[width=\linewidth]{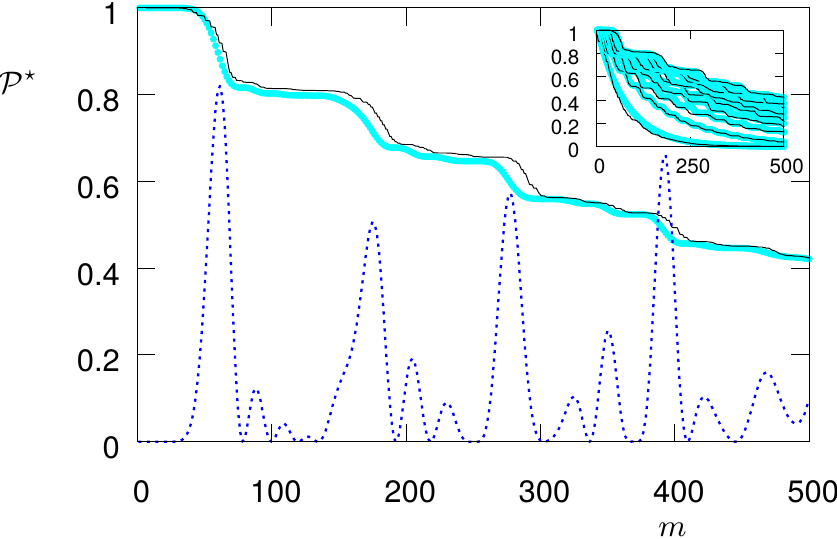}
\caption{\textbf{Left-most qubit excited.} $\mathcal{P}^\star$ as a function of the number of measurements $m$ as compared to Eq.~\eqref{eq:P-average-exc} for $\lambda=9$. The blue dashed line is $|c_{9}(t)|^2$ as obtained by a simulation with $H_{9}$, while in the inset $\mathcal{P}^{\star}$ for $\lambda=1,\dots,9$ (from bottom to top) is shown. The probability density function is bimodal with $p_1=p_2=0.5$, $\mu^{(1)}=1\,\mathrm{\mu s}$, and $\mu^{(2)}=5\,\mathrm{\mu s}$.}
\label{fig:fe-P-average-exc}
\end{figure}
For $\lambda=2,\dots,10$ we determine the time when the excitation first peaks at the edge qubit $\lambda$ by evaluating the numerical simulations. This allows us to determine the velocity by a fit, yielding $v\approx0.06\,$sites$/$ms compared to a theoretical bound given by the norm of the interaction operator \cite{Kliesch}, i.e. $e\Vert\beta(\sigma_{x}^{\lambda}\sigma_{x}^{\lambda + 1}+\sigma_{y}^{\lambda}\sigma_{y}^{\lambda + 1})/2\Vert$ yielding $\approx 0.085\,$sites$/$ms.

Fig.~\ref{fig:fe-P-average-exc} shows the survival probability as obtained by a numerical simulation (black) compared to Eq.~\eqref{eq:P-average-exc} (cyan) for $\lambda = 9$ (in the inset the most probable value $\mathcal{P}^{\star}$ is shown for $\lambda = 1,\dots,9$, bottom to top). The plateaus correspond to zero or very little excitation of the edge qubit ($|c_{\lambda}|$ very small), while the steps correspond to considerable excitation located at the edge qubit. The remnant plateaus for $\lambda=1$ occur only in the numerical simulation and are absent in the model since they do not come from an oscillation of the excitation in the 1-qubit subspace, but instead from repetitive measurements after the smaller time interval $\mu^{(1)}$, i.e. an effect that is averaged out in the model.

\subsubsection{Coherent Couplings}
Here, we compare the results obtained above to those obtained via coherent coupling. The coherent coupling is included by considering the additional coupling Hamiltonian $H_c(\lambda)=\left(\sigma_x^{\lambda+1}\sigma_x^{\lambda+2} + \sigma_y^{\lambda+1}\sigma_y^{\lambda+2} \right)$. We choose the coupling such that $g=\pi/(2\overline{\mu})$ in the case of continuous coupling, and $s=\pi/2$ in the case of pulsed coupling. Thus, on average in both cases the pulse area of the coupling is the same, and for the pulsed coupling the projective measurement is substituted by an excitation flip between qubits $\lambda+1$ and $\lambda+2$.
\begin{figure}[t]
\centering
 \includegraphics[width=\linewidth]{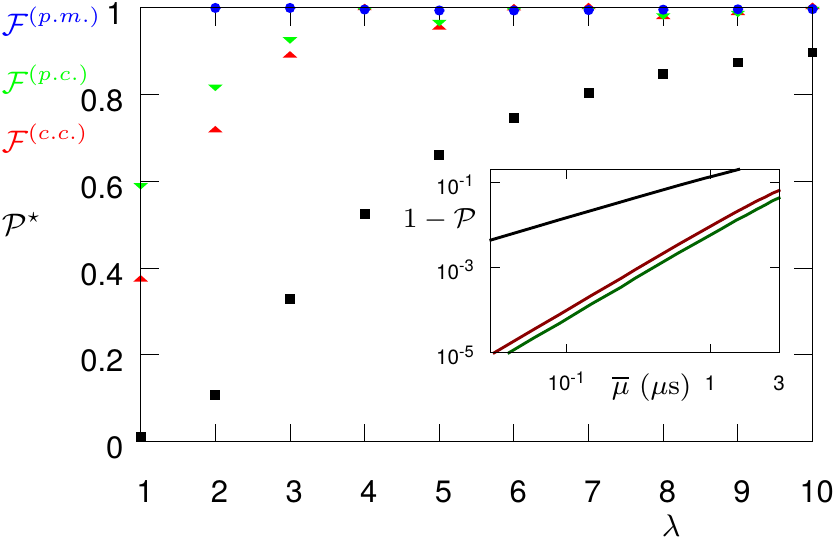}
 \caption{\textbf{Performance of the Zeno protocols as a function of the subspace size $\lambda$.} The red upper triangles, green lower triangles and blue circles show the fidelities, respectively, for continuous coupling, pulsed coupling and projective measurements. Instead, the black squares show the survival probability. The simulations where carried out for the initial W-state and a bimodal probability density function with $p_1 = p_2 = 0.5$, $\mu^{(1)}=3\,\mathrm{\mu s}$, and $\mu^{(2)}=5\,\mathrm{\mu s}$. The inset shows how the system behaves for $\lambda=5$ when $m\overline{\mu}$ is constant, and the interaction (given by the number of measurments $m$ or the coherent coupling strength $g$) is varied: As we approach the Zeno limit the confinement error $1-\mathcal{P}$ vanishes for all the three Zeno protocols (from top to bottom: p.m (black), c.c. (dark red), p.c. (dark green)), and the scaling with respect to $\overline{\mu}$ is linear for the protocol based on projective measurements and quadratic for the coherent coupling methods.}
 \label{fig:lambda-methods}
\end{figure}
Fig.~\ref{fig:lambda-methods} shows the fidelity $\mathcal{F}$ (see appendix for details) of the respective dynamics as a function of the number of qubits $\lambda$ composing the subspace. While projective measurements $(p.m.)$ yield the highest fidelity, all three Zeno protocols show a similar scaling behaviour with respect to $m$ and $\lambda$. It should be noted though that due to the probabilistic nature of the projective measurements given by the survival probability $\mathcal{P}^{\star}$, the coherent methods show the better deterministic performance with a slight advantage for pulsed coupling $(p.c.)$ over coherent coupling $(c.c.)$. For increasing $\lambda$ we approach higher values of fidelity and survival probability, since the edge qubit is on average less populated and we have less leakage. The inset of Fig.~\ref{fig:lambda-methods} shows the leakage $1-\mathcal{P}$ for the three protocols, projective measurements (black), pulsed coupling (dark green) and continuous coupling (dark red), when approaching the Zeno limit: we set $m\overline{\mu}$ to be a constant value and we decrease $\overline{\mu}$ while at the same time $m$ is increasing. The results are for $\lambda=5$, $p_1=1$, and $\mu^{(1)}=3\,\mathrm{\mu s}$. While the projective measurements approach shows a linear scaling with $\overline{\mu}\propto 1/m$, the coherent coupling protocols exhibit a quadratic scaling (see the inset of Fig.~\ref{fig:lambda-methods}). The linear scaling in the first case is a direct consequence of Eq.~(\ref{eq:P-strictZeno3}), while the quadratic scaling in the latter case corresponds to the prediction of the off-resonant driving model, as shown above.
\begin{figure}[t]
\centering
 \includegraphics[width=\linewidth]{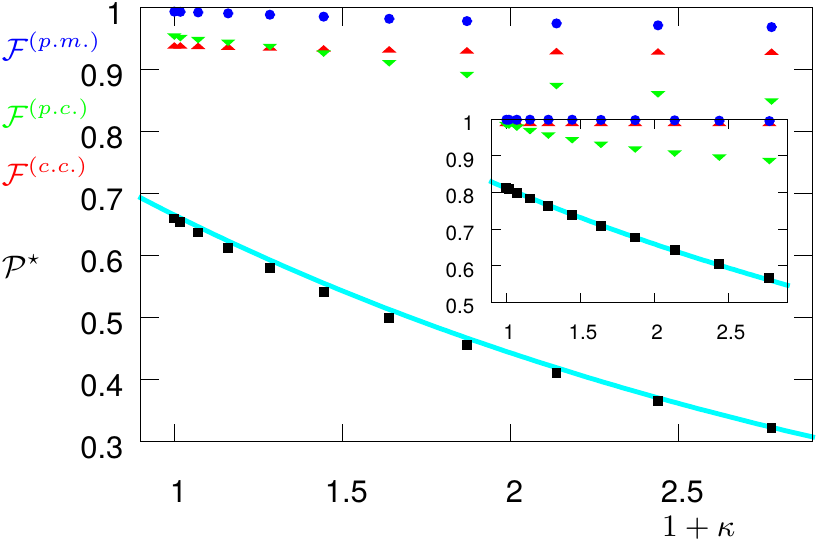}
 \caption{\textbf{Performance of the three protocols as a function of the time disorder $1 + \kappa$.} The red upper triangles, green lower triangles and blue circles show the fidelities, respectively, for continuous coupling, pulsed coupling and projective measurements. The black squares show the survival probability. The system was initially prepared in the W-state. The cyan curve is the theoretical value obtained by Eq.~\eqref{eq:P-average-exc}, where $|c_{\lambda}(t)|$ has been taken from the time evolution with $H_{\lambda}$. The probability density function is bimodal with $p_1=0.8$, $p_2=0.2$, $\overline{\mu}=3\,\mathrm{\mu s}$, $\mu^{(1)}\in[1,3]\,\mathrm{\mu s}$, and $\mu^{(2)}\in[3,11]\,\mathrm{\mu s}$, corresponding to $\kappa\in[0,1.778]$. Inset: Initially only the left-most spin was excited.}
 \label{fig:kappa-methods}
\end{figure}
Finally, in Fig.~\ref{fig:kappa-methods} the performance of the Zeno protocols is shown as a function of the time disorder $1 + \kappa$. As it can be observed, we find a decrease in the fidelity $\mathcal{F}$ both for the protocol based on projective measurement (p.m.), and for the coherent pulsed coupling (p.c.), while, trivially, no change occurs for continuous coupling $(c.c.)$.

\section{Discussion and Outlook}
We have investigated how the presence of a temporal stochasticity in the observation of a quantum system can perturb its dynamics, also in an irreversible manner. In particular, we have studied the accessibility to quantum Zeno dynamics when stochasticity in the quantum Zeno protocols is taken into account, and we have shown that the system dynamics is confined to a subspace when a strong system-environment interaction is switched on at random (frequent enough) times. This work extends the large deviation theory approach to stochastic quantum Zeno phenomena~\cite{Gherardini1}, to the description of survival probabilities in quantum Zeno dynamics~\cite{Smerzi1}.
On top of that the new approach has allowed us to introduce a weak Zeno regime, where the observation is frequent enough to ensure that the dynamics in the subspace follow closely the dynamics of a perfectly truncated system but the survival probability decays with increasing time.
Besides the stochastic quantum Zeno protocol based on projective measurements, we have shown that stochastic Zeno phenomena can be equivalently achieved with high fidelity $\mathcal{F}$ by applying fast random unitary kicks or strong continuous couplings. In particular, we have shown that a possible experimental realisation of this dynamical regime can be achieved by means of a continuous coupling, that has the advantage to be fully deterministic and easy to implement. If (almost) perfect confinement is required, it can be realised by projective measurements at the price of a probabilistic protocol.
Only by modelling with enough accuracy the nature of such interactions with the environment, quantum dynamics might be effectively controlled in well-defined Hilbert space portions. Apart from confining the state of the system to a static Zeno subspace, this is relevant as well for the transfer of arbitrary physical states between decoherence-free subspaces \cite{Lidar1}, enabled by engineered protocols for Zeno-protection \cite{Paz-Silva1}.
To conclude, our formalism is able to take into account external noise sources affecting the quantum system, and it can be applied within the general context of an open quantum system \cite{Nielsen1} in interaction with an external environment, whose evolution is well-described by a trace preserving and completely positive (CTCP) map, also called quantum channel \cite{Caruso1}.
The approach is completely independent of the platform and possible platforms include trapped ions \cite{Blatt}, neutral atoms \cite{Schafer1,Signoles1,Bloch2008}, quantum dots \cite{Hanson} and superconducting qubits \cite{Niskanen}.
The results are expected to move further steps towards the development and implementation of Zeno enabled quantum technology.


\appendix
\section{Appendix}
\subsection{Weak and strong Zeno regime}
We want to find an approximation for the survival probabilities most probable value $\mathcal{P}^\star$ as given by Eq.~(\ref{eq:P_geom}) when the confinement is good but not perfect. To this scope we start by a Taylor expansion of $\ln q(\mu)$ as a function of the time interval times: we define $\alpha_{k}\equiv\left.\frac{\partial^{k}\ln(q(\mu))}{\partial\mu^{k}}\right|_{\mu = 0}$ and write
\bea\label{eq:Pstar-expansion}
\mathcal{P}^{\star} = \exp\left\{m\sum_{k = 1}^{\infty}\frac{\alpha_{k}}{k!}\int_{\mu}d\mu p(\mu)\mu^{k}\right\}
\nonumber\\
= \exp\left\{m\sum_{k = 1}^{h/2}\frac{\alpha_{2k}}{2k!}\int_{\mu}d\mu p(\mu)\mu^{2k} + R_{h}(\xi)\right\},
\eea
where $R_{h}(\xi)$ is the remainder of Taylor expansion of $\ln(q(\mu))$ up to the $h-$th order, where $\xi\in[0,\mu]$ is a real number. For odd $k$ due to the symmetry of $q(\mu)$ we find $\alpha_{k} = 0$. Thus $h$ is assumed to be an even number, greater than zero. For $h = 2$, namely by considering a second order approximation of the Taylor expansion (only the first term of the summation in Eq.~\eqref{eq:Pstar-expansion} is considered), the survival probability's most probable value is equal to $\mathcal{P}^{\star} = \exp\left\{m\frac{\alpha_{2}}{2}(1+\kappa)\overline{\mu}^2\right\}\exp\left\{m\langle R_{2}(\xi)\rangle\right\}$, where $\alpha_{2} = -2\Delta^{2}_{\rho_{0}}H_{\Pi}$, $\kappa=\frac{\Delta^{2}\mu}{\overline{\mu}^2}$ and $\overline{\mu}$ and $\Delta^{2}\mu$ are, respectively, the expectation value and variance of $p(\mu)$. The $2$nd order remainder of the Taylor expansion in the Lagrange form is $R_{2}(\xi)\equiv\frac{\partial^{3}\ln(q(\mu))}{\partial\mu^{3}}\big|_{\mu = \xi}\frac{\mu^{3}}{6}$, where $\left|\frac{1}{6}\frac{\partial^{3}\ln(q(\mu))}{\partial\mu^{3}}\big|_{\mu = \xi}\right|\leq C$ for some positive constant $C$ that depends on the form of the specific system Hamiltonian $H$ and the initial state $\rho_{0}$. Hence, $\langle R_{2}(\xi)\rangle\equiv\int_{\mu}d\mu p(\mu)R_{2}(\xi)$ is bounded by $C\mu^3$ and, if
\be
\int_{\mu}d\mu p(\mu)\mu^{3}\ll\frac{1}{mC}
\ee
(which is Eq.~(\ref{eq:skewness_condition})),
the term $\langle R_{2}(\xi)\rangle$ is negligible, such that
\be
\mathcal{P}^\star\approx
\exp\left\{-m\Delta^2_{\rho_{0}}H_{\Pi} (1+\kappa)\overline{\mu}^2 \right\}
\ee
which is Eq.~(\ref{eq:P_zeno2}).
It is worth to note that the validity of Eq.~\eqref{eq:P_zeno2} (i.e. the condition Eq.~\eqref{eq:skewness_condition}) does not depend on the variance of the probability distribution $p(\mu)$, but on its degree of skewness.

\subsection{Fidelity for quantum Zeno dynamics}
The performance of the Zeno protocols for Zeno dynamics can be evaluated by introducing the Uhlmann fidelity \cite{Jozsa,Uhlmann}
\be\label{eq:fidelity}
\mathcal{F}^{(protocol)}=\text{Tr}\sqrt{\sqrt{\rho_m^{(\Pi)}} \rho_m^{(protocol)} \sqrt{\rho_m^{(\Pi)}}},
\ee
which compares the evolved density matrices to the density matrix $\rho_{m}^{(\Pi)}\equiv\rho^{(\Pi)}(t = t_{m})$ obtained by exact subspace evolutions. Finally,  \textit{(protocol)} refers to the examined Zeno protocols projective measurements $(p.m.)$, continuous coupling $(c.c.)$ or pulsed coupling $(p.c.)$.

\begin{acknowledgements}
We acknowledge S. Ruffo, A. Smerzi and F.S. Cataliotti for useful discussions. This work was financially supported by the Ente Cassa di Risparmio di Firenze through the project Q-BIOSCAN.\\
\quad\\
{\bf Author contributions.} M.M. and S.G. contributed equally to this work. M.M. and S.G. carried out the analytic calculations. M.M. performed the numerical simulations. F.C. conceived the project, led the theory and supervised the work. All authors contributed to the discussion, analysis of the results and the writing of the manuscript.
\end{acknowledgements}

\end{document}